\documentclass[twocolumn,appendixfloats]{aastex6_1}
\usepackage{graphicx}
\usepackage{natbib}
\usepackage{latexsym}
\usepackage{amssymb}
\usepackage{longtable}
\usepackage{amsmath}
\usepackage{url}
\def\msun{\ifmmode {\rm\,M_\odot}\else ${\rm\,M_\odot}$\fi}
\def\Msun{\ifmmode {\rm\,\it{M_\odot}}\else ${\rm\,M_\odot}$\fi}
\def\lsun{\ifmmode {\rm\,L_\odot}\else ${\rm\,L_\odot}$\fi}
\def\Lsun{\ifmmode {\rm\,\it{L_\odot}}\else ${\rm\,L_\odot}$\fi}
\def\rsun{\ifmmode {\rm\,R_\odot}\else ${\rm\,R_\odot}$\fi}
\def\Rsun{\ifmmode {\rm\,\it{R_\odot}}\else ${\rm\,R_\odot}$\fi}
\def\Tsun{\ifmmode {\rm\,T_\odot}\else ${\rm\,T_\odot}$\fi}
\def\arcsec{\ifmmode {^{\prime\prime}}\else $^{\prime\prime}$\fi}
\def\asec{\ifmmode {^{\prime\prime}}\else $^{\prime\prime}$\fi}
\def\arcmin{\ifmmode {^{\prime}}\else $^{\prime}$\fi}
\def\amin{\ifmmode {^{\prime}}\else $^{\prime}$\fi}
\def\simlt{\mathrel{\spose{\lower 3pt\hbox{$\mathchar"218$}}
     \raise 2.0pt\hbox{$\mathchar"13C$}}}
\def\simgt{\mathrel{\spose{\lower 3pt\hbox{$\mathchar"218$}}
\     \raise 2.0pt\hbox{$\mathchar"13E$}}}

\begin{document}

%\title{The dramatic evolution of circumstellar absorption \\
%lines associated with WD 1145+017}

\title{Evidence for eccentric, precessing gaseous debris in the \\ circumstellar absorption towards WD 1145+017}

\author{P. Wilson Cauley}
\affiliation{Astronomy Department, Wesleyan University, Van Vleck Observatory, Middletown, CT}
\affiliation{School of Earth and Space Exploration, Arizona State University, Tempe, AZ}

\author{Jay Farihi}
\affiliation{Department of Physics and Astronomy, University College London, London, UK}

\author{Seth Redfield}
\affiliation{Astronomy Department, Wesleyan University, Van Vleck Observatory, Middletown, CT}

\author{Stephanie Bachmann}
\affiliation{Department of Physics, Indiana University of Pennsylvania, Indiana, PA}
\affiliation{Astronomy Department, Wesleyan University, Van Vleck Observatory, Middletown, CT}

\author{Steven G. Parsons}
\affiliation{Department of Physics and Astronomy, University of Sheffield, Sheffield, UK}

\author{Boris T. G\"{a}nsicke}
\affiliation{Department of Physics, University of Warwick, Coventry, UK}

\correspondingauthor{P. Wilson Cauley}
\email{pwcauley@gmail.com}

\begin{abstract} 

We present time-series spectra revealing changes in the circumstellar line 
profiles for the white dwarf
WD 1145+017. Over the course of 2.2 years, the spectra show complete velocity reversals in the
circumstellar absorption, moving from strongly red-shifted in 2015 April to strongly
blue-shifted in 2017 June. The depth of the absorption also varies, 
increasing by a factor of two over the same period. The dramatic 
changes in the line profiles are consistent with eccentric circumstellar 
gas rings undergoing general relativistic precession. As the argument of periapsis of 
the rings change relative to the line of sight, the transiting gas shifts from 
receding in 2016 to approaching in 2017. Based on the precession
timescales in the favored model, we make predictions for the line 
profiles over the next few years. Spectroscopic monitoring of WD 1145+017 will test 
these projections and aid in developing more accurate white dwarf accretion disk models. 

\end{abstract}


\section{INTRODUCTION}
\label{sec:intro}

The polluted atmospheres observed in at least $\approx 30\%$ of white dwarfs are caused
by the accretion of planetesimal debris, the result of tidal disruption of rocky bodies by the host star
and the eventual collisional grinding of the debris into dust 
\citep{zuckerman10,koester14,kenyon17a}. In some systems, the disrupted planetesimal material is detected in
disks of dust and gas \citep[see][and references therein]{farihi16}. 
The atmospheric abundances of heavy elements reflects the composition of the orbiting and 
infalling planetesimal debris, and hence the bulk composition of the disrupted parent body. Polluted
white dwarf atmospheres thus offer an exciting window into the chemistry and assembly
of the planets and minor bodies formed in the system \citep{gansicke12,jura14,raddi15,farihi13,xu13}. 

WD 1145+017 was the first system for which transits associated with closely orbiting planetesimals were
detected: \citet{vanderburg15} presented \textit{K2} light curves consistent with large
clouds of opaque debris orbiting near the stellar Roche limit. Subsequent photometric monitoring
has revealed significant evolution of the transit depths and morphologies
\citep{rappaport16,gansicke16,croll17}. \citet{xu16} presented high-resolution spectra of WD 1145+017
that showed circumstellar gas lines were present in several metal species 
with depths $\approx 10 - 30\%$ below the continuum. The red-shifted line components exhibited 
maximum velocities of $\approx 200$ km s$^{-1}$, and significant absorption extending down
to $\approx 0$ km s$^{-1}$, indicative of material receding along the 
line-of-sight across a wide range of velocities. \citet{redfield17} presented multi-epoch 
spectra that displayed short- and long-term variability of the circumstellar absorption,
and suggested an eccentric circumstellar disk model to account for the red-shifted velocities. 

In this paper we present a multi-year monitoring campaign of
the circumstellar absorption towards WD 1145+017. The data reveal that the absorption velocities 
undergo a gradual change from strongly red-shifted to strongly blue-shifted over the
course of $2.2$ years. The change from red-shifted to blue-shifted 
velocities is evidence that the 
absorbing gas is not infalling onto the star. We model the line profiles using eccentric
rings that are precessing on a timescale similar to what is predicted by general relativity
for the mean periastra of the rings,
shifting the line-of-sight absorbing
material from moving away from us to towards us with a precession period of $\approx 5.3$ 
years. We derive circumstellar line profile predictions based on the ring model to be 
tested with future spectroscopic monitoring. In \autoref{sec:observations} we briefly
describe the observations and data reduction steps. \autoref{sec:model} details the 
eccentric ring model and \autoref{sec:discussion} is a discussion of the model and
other possible explanations for the velocity changes in the line profiles. 
Finally, a brief summary and our conclusions are give in \autoref{sec:conclusion}.   

\section{OBSERVATIONS AND DATA REDUCTION}
\label{sec:observations}

The data were obtained using two different instruments: HIRES on Keck I \citep{vogt94} and X-shooter on the
VLT \citep{vernet11}. The dates and details of the observations are given in
\autoref{tab:observations}. Further information concerning these data can be 
found in \citet{xu16} and \citet{redfield17}, respectively. 

The 2016 and 2017 Keck data sets were collected using the same instrument settings
as the other Keck observations. The reduced 2015 April 11 data
was previously published \citep{xu16}. All 2016 and 2017 Keck data
were reduced using the HIRES Redux package\footnote{Written by Jason X. 
Prochaska; http://www.ucolick.org/~xavier/HIRedux/index.html} following typical
reduction steps.
The 2017 February 14 X-shooter data were obtained with the same settings as the other
X-shooter data and reduced in an identical manner.   

%which we briefly summarize here. The spectra were extracted using version 2.6.8
%of the X-shooter Common Pipeline Library. Standard recipes were used to perform the array
%calibration and for deriving wavelength solutions. Each spectrum was optimally extracted assuming
%the observation was taken in staring mode.  

\begin{deluxetable*}{lcccccccc}
\tablecaption{Observations \label{tab:observations}}
\tablehead{\colhead{Date}&\colhead{}&\colhead{}&\colhead{}&\colhead{$t_\text{exp}^a$ }&\colhead{S/N}&\colhead{}&\colhead{}&\colhead{Spectral range}\\
\colhead{(UT)}&\colhead{Instrument}&\colhead{Program ID}&\colhead{$N_\text{exp}$}&\colhead{(s)}&
\colhead{(@ 5000 \AA)}&\colhead{Slit width$^a$}&\colhead{\textit{$\lambda / \Delta \lambda$}}&\colhead{(\AA)}}
\colnumbers
\tabletypesize{\scriptsize}
\startdata
2015 Apr 11 & HIRES & U027Hb & 3 & 2400 & 25 & 1.$\!\arcsec$148 &37\,000 & 3100 - 5950 \\
2015 Nov 14 & HIRES & N116Hb & 2 & 1200 & 10 & 1.$\!\arcsec$148 &37\,000 & 3100 - 5950 \\
2016 Feb 14 & X-shooter & 296.C-5014(A) & 40 & 300 & 120 & 0.$\!\arcsec$9 &7000 & 3000 - 10\,000\\
2016 Mar 29 & X-shooter & 296.C-5014(D) & 29 & 280 & 130 & 0.$\!\arcsec$9 &7000 & 3000 - 10\,000\\
2016 Apr 8 & X-shooter & 296.C-5014(B) & 44 & 280 & 140 & 0.$\!\arcsec$9 &7000 & 3000 - 10\,000 \\
2016 Jun 3 & HIRES & N171Hb & 3 & 1200 & 13 & 1.$\!\arcsec$148 & 37\,000 & 3500 - 7400 \\
2017 Feb 14 & X-shooter & 598.C-0695(A) & 48 & 280 & 120 & 0.$\!\arcsec$9 & 7000 & 3000 - 10\,000 \\
2017 May 17 & HIRES & N188Hb & 3 & 1800 & 16 & 1.$\!\arcsec$148 & 37\,000 & 3500 - 7400 \\
2017 Jun 26 & HIRES & N188Hb & 3 & 1200 & 18 & 1.$\!\arcsec$148 & 37\,000 & 3500 - 7400 \\
\enddata
\tablenotetext{a}{X-shooter exposure times and slit widths are listed for the UVB arm}
\end{deluxetable*}

Photospheric models are employed to remove the narrow atmospheric metal 
lines \citep[see][for details]{koester10,redfield17}.
The photospheric absorption is present at a velocity of $\approx 42$ km s$^{-1}$. We assume
a mass of 0.6 M$_\text{Sun}$ which produces a gravitational red-shift of 
30 km s$^{-1}$ and thus a system radial velocity of
12 km s$^{-1}$. Due to the large widths and maximum absorption
velocities of the circumstellar line profiles, choosing a different value of the gravitational red-shift
does not change the interpretation of the line profiles in a 
meaningful way. Once the photospheric model is subtracted from the spectrum, 
the resulting circumstellar profile is shifted into the observer's reference frame.
We note that no detectable variability is present in the photospheric metal lines. 

%For example, assuming $M_* = 0.8 M_\odot$ gives a red-shift of $\approx 50$
%km s$^{-1}$

\begin{figure*}[ht!]
   \centering
   \includegraphics[scale=.85,clip,trim=10mm 20mm 5mm 5mm,angle=0]{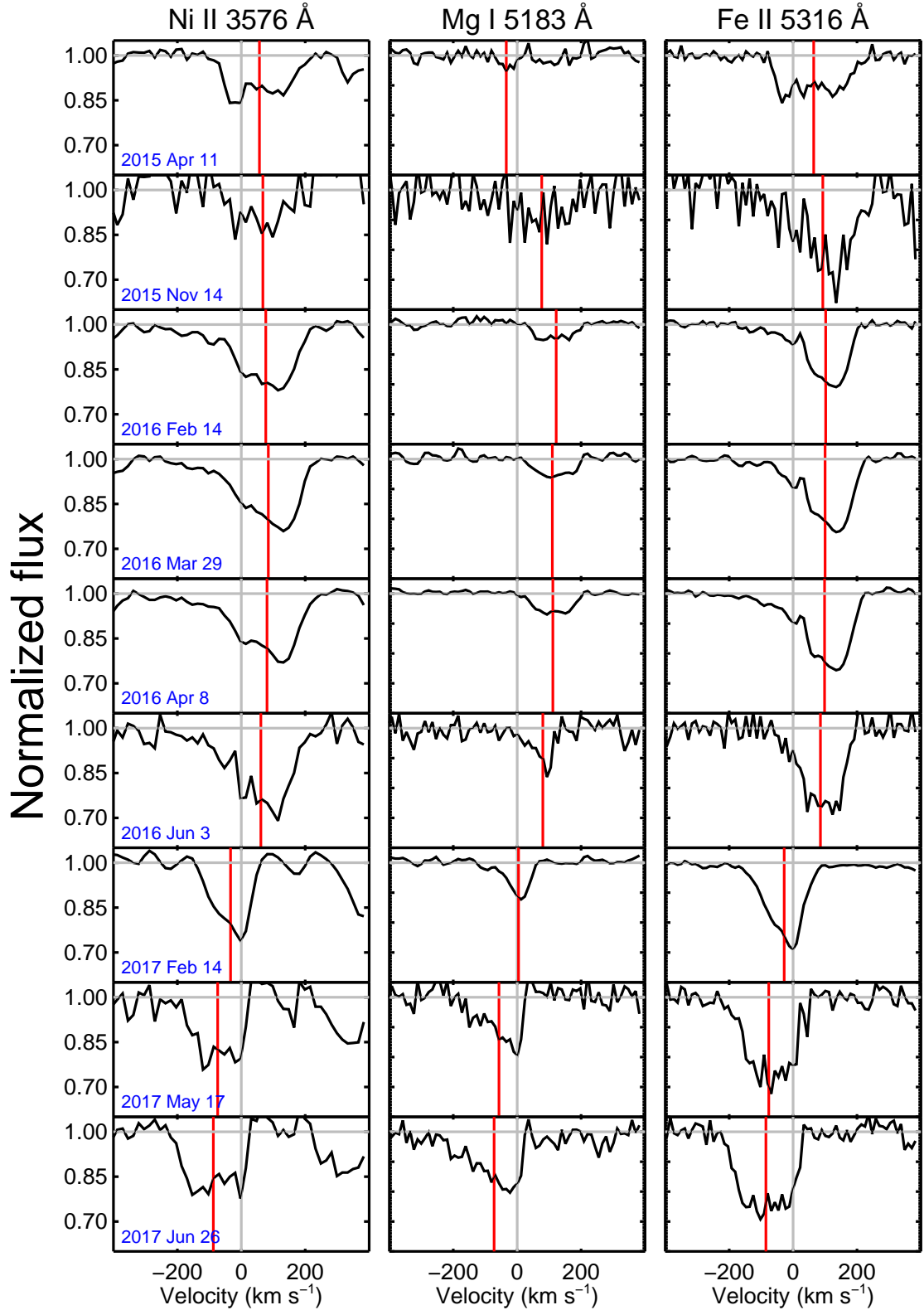}
   \figcaption{Circumstellar line profiles for the \ion{Ni}{2} 3576 \AA, \ion{Mg}{1} 5183 \AA,
   and \ion{Fe}{2} 5316 \AA\ lines. All profiles are binned to the resolution of the 
   X-shooter data. The normalized flux continuum and zero velocity are marked with
   horizontal and vertical gray lines, respectively. The photospheric absorption
   lines have been subtracted and all profiles are shown in the rest frame of the observer. The vertical red lines show the absorption-weighted velocity of the circumstellar line.
   The lines show a common evolution in terms of strength and morphology. Note
   the appearance of the longer wavelength lines in the \ion{Ni}{2} profiles 
   when the absorption changes from red-shifted to blue-shifted.
    \label{fig:profs}}
\end{figure*}

The reduced circumstellar line profiles for \ion{Ni}{2} 3576.764 \AA, \ion{Mg}{1} 5183.604 \AA,
and \ion{Fe}{2} 5316.615 \AA\ are shown in \autoref{fig:profs}. All Keck profiles
are binned to the X-shooter resolution. Due to the $\approx 150-200$ km s$^{-1}$ velocity widths of the
circumstellar absorption, little information is lost in the binning. All three lines show
similar evolution in time, moving from absorption with maximum red-shifted velocities of
$\approx 200$ km s$^{-1}$ in 2015 to maximum blue-shifted velocities of $\approx 200$ km s$^{-1}$
in 2017. This suggests that all three metal ions trace coincident regions
of the circumstellar material. In the next section we present a model that accounts
for the absorption via eccentric gas rings that are precessing due to general
relativistic effects of the white dwarf's gravity well.

\bigskip

\section{PRECESSING ECCENTRIC RINGS}
\label{sec:model}

\begin{figure}[htbp]
   \centering
   \includegraphics[scale=.35,clip,trim=25mm 30mm 10mm 35mm,angle=0]{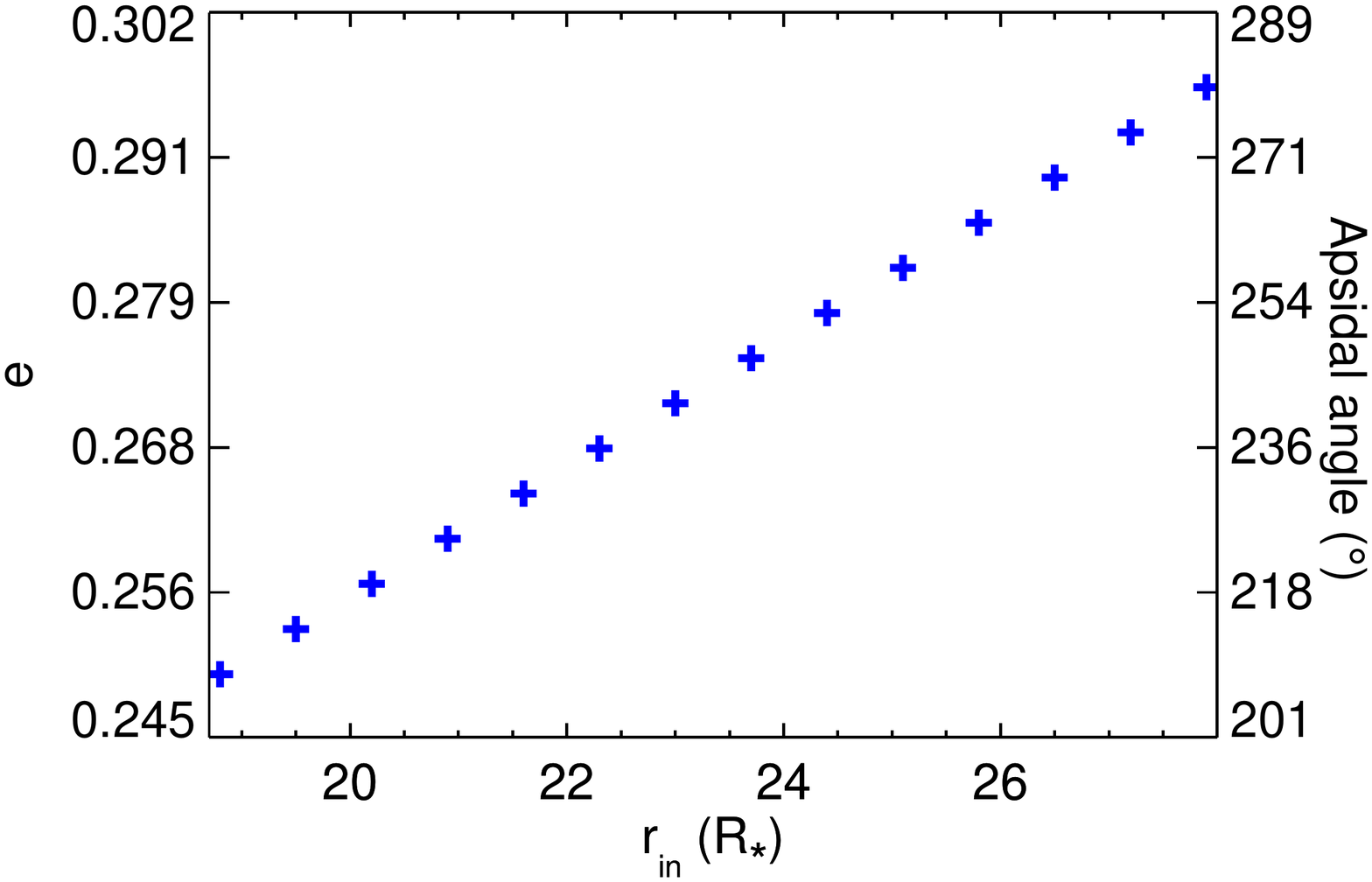}
   \figcaption{Periastron versus eccentricity and apsidal angle. The
   apsidal line angles of the rings are misaligned and the rings
   become more eccentric with larger $r_{\text{in}}$. \label{fig:rinphi}}
\end{figure}

We model the absorption using fourteen confocal, eccentric gas rings in co-planar orbits oriented
edge-on to the line-of-sight. The rings are parameterized according to their periastra, eccentricity,
and the position angle of the apsidal line measured clockwise from the line-of-sight
direction \citep[see Appendix D of][]{metzger12}, i.e., the apsidal lines of the
rings are misaligned. It is also assumed that the rings orbit at their Keplerian
 velocities, ignoring pressure support that slightly reduces gas velocity 
\citep[e.g.,][]{rafikov11}. Each ring has a radial width of 0.5 $R_*$ and a
constant azimuthal density. A density is assigned to the innermost ring, which 
has periastron distance $r_{\text{in}}$, for each epoch and then
densities in subsequent rings orbiting at distance $r$ are scaled by a factor of $(r_{\text{in}}/r)^2$.
The ring width is arbitrary and is degenerate with the ring density, which combine to 
give the optical depth through the gas.  
The eccentricity of the rings changes linearly from 0.25 to 0.30 for periastron distances 
of 18.8 $R_*$ to 27.9 $R_*$ and the apsidal line angle steps linearly from 205$^\circ$ for
the innermost ring to 283$^\circ$ for the outermost ring. The ring eccentricities and apsidal
line angles are shown as a function of $r_{\text{in}}$ in \autoref{fig:rinphi}. 
The precession period of the rings is 5.3 years, which is chosen in order to match the observed 
 profile evolution. This is very similar to the mean precession period of 6.3 years given by general 
 relativity for the periastron ring distances in our model, although we note that general relativistic
  precession is a strong function of periastron distance \citep{veras14,manser16}.  

The line profiles are generated using Doppler-broadened delta functions. We take
the Doppler width of the line profile to be the thermal line width at that point in the ring.
For the case of \ion{Fe}{2}, the thermal line widths are $\approx 2$ km
s$^{-1}$ at gas temperatures of $\approx 6000$ K, which is typical for the orbital distances
of the ring material \citep{melis10}. The scale height, which determines the density as a function of
vertical distance from the ring mid-plane, at each ring position is calculated using
the same temperature approximation. The stellar surface is divided into a $100 \times 100$
Cartesian grid and the intensity from each stellar grid point is then extincted by any obscuring 
ring material. We do not consider multiple scatterings through the rings. The entire stellar 
grid is then summed to produced the final line profile. Limb darkening is accounted for using a
4-parameter Claret coefficient law for $T_{\rm eff} = 16\,000$ K and log $g = 8.0$ \citep{gianninas13}. 

It is important to note that the ring configuration described here is almost certainly not unique in being
able to broadly reproduce the observed line profiles. Furthermore, the WD 1145+017 system is highly dynamic
and new material appears to be constantly replenished through the disk. However, there are several
critical constraints on the eccentric ring model. First, the combination of eccentricity, viewing 
angle, and periastron distance must combine to produce orbiting gas at velocities of 
$\pm$200 km s$^{-1}$ along the line-of-sight. The periastra of the rings must also have general 
relativistic precession periods similar to
that observed. Values of the eccentricity $< 0.1$ cannot produce sufficient line-of-sight
velocities unless the orbital distances are $\lessapprox 10 R_*$. 
However, a ring with this smaller periastron
distance results in a precession period of days to months, shorter than suggested by the changing line
profiles. Alternatively, periastra $\gtrapprox 50 R_*$ require eccentricities $> 0.4$ 
to produce the observed velocities in the absorption profiles. In this case, the predicted 
general relativistic precession periods are $\approx 20-30$ years. Thus, assuming that a 
complex of precessing gas rings is the correct geometry, the actual configuration should exist
within this range. 

\begin{figure}[htbp]
   \centering
   \includegraphics[scale=.7,clip,trim=30mm 20mm 65mm 0mm,angle=0]{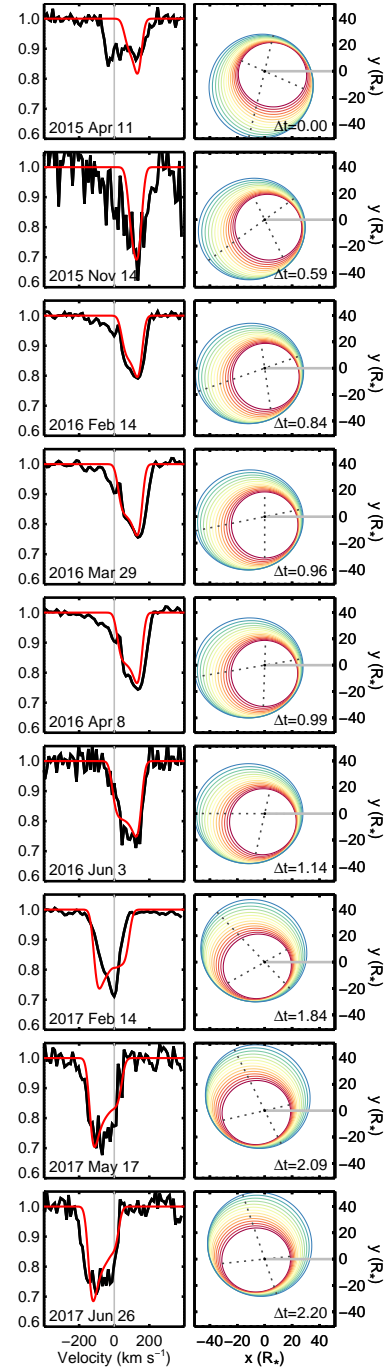}
   \figcaption{The \ion{Fe}{2} 5316 \AA\ line profiles in the observer rest frame 
   (black) and the corresponding eccentric disk model profiles (red) are shown 
   in the left-hand panels. The right-hand
   panels show the configuration of gas rings at each epoch. The time in years from the 2015 April 11
   observation is given by $\Delta t$. The colors change from red
   for the innermost and least eccentric rings to green for the outermost and most
   eccentric rings. The solid gray line indicates the line-of-sight to Earth and
   the dotted charcoal lines show the apsidal directions for the inner and outermost
   rings. Note that the apsidal directions are misaligned for the rings,
   a condition that is necessary in the model to reproduce the $\approx 200$ km s$^{-1}$
   line widths. \label{fig:diskmod}}
\end{figure}

The model line profiles are shown overlaid on the \ion{Fe}{2} 5316 \AA\ data in
\autoref{fig:diskmod}. In general, the precessing rings follows the evolution of the
observed line profiles. The major exception is the model profile from 2015 April 11,
which is weaker and shows a distinct absorption component near zero in
\autoref{fig:profs}. There may be additional
absorbing gas components that are responsible for the deviations 
from the ring profiles modeled here, 
which is plausible in the apparently dynamically active environment of this system.
Thus, we have not attempted to accurately represent the entirety of absorbing gas
during the 2015 April epoch. Furthermore, we do not attempt to model the zero velocity
absorption component, which also seems to be present in the other line profiles at most epochs, 
since it appears to be fairly static. This component could be 
due to a stable outer gas ring with low eccentricity or a viewing angle that 
intersects a low line-of-sight velocity. 

\begin{figure}[htbp]
   \centering
   \includegraphics[scale=.38,clip,trim=25mm 25mm 25mm 20mm,angle=0]{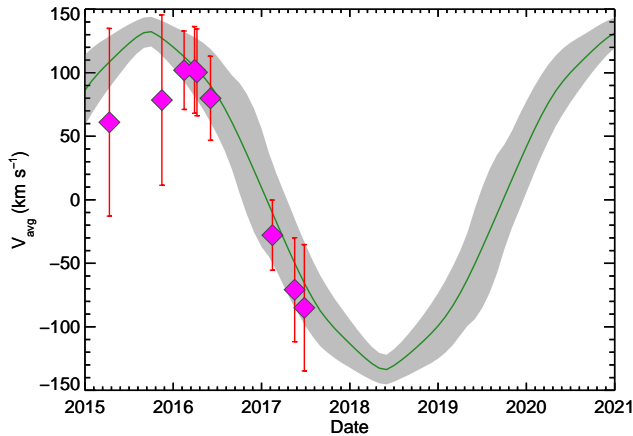}
   \figcaption{Absorption-weighted average velocities for the \ion{Fe}{2} 5316 \AA\ line
   profiles (pink diamonds) and the corresponding model values from the eccentric disk profiles. 
   The uncertainties (red lines and gray band) are calculated by taking the 
   standard deviation of the velocities 
   that comprise 68\% of the measured absorption. The model velocities and line widths
   match the trend of the observed line profiles, although the 2015 lines are much
   broader than the model profiles. The model predicts that the 
   profiles will switch back to being dominated by
   red-shifted absorption some time in mid-2019. \label{fig:vels}}
\end{figure}

\autoref{fig:vels} shows the absorption-weighted velocities of the observed \ion{Fe}{2} 5316 \AA\
line profiles (pink diamonds) and the model line profiles (green line). The 
velocities are projected out to 2021 when the rings will have undergone a full
precession period. The red error bars in \autoref{fig:vels} are calculated by taking the
standard deviation of the velocities that comprise 68\% of the total absorption. The
same calculation is performed for the model line profiles and this is shown
with the gray band surrounding the model $V_{avg}$ values. Thus
the larger uncertainties simply represent broader, flatter profiles.
With the exception of the 2015 line profiles, the observed $V_{avg}$ values and 
line widths are in good agreement. 
The model predicts that the absorption will switch 
back to red-shifted velocities sometime in mid-2019. Continued monitoring of WD 1145+017 
over the next two years will allow the precessing ring model to be robustly tested.

%Note that the velocity curve in
%2020 does not have the same amplitude as in 2015. This is due to the different precession
%periods associated with the different rings, creating profile shapes that do not repeat
%at a single frequency. Continued monitoring of WD 1145+017 over the next two years will 
%allow the precessing disk model to be robustly tested.  

\section{Discussion}
\label{sec:discussion}

The eccentric and precessing disk model discussed above can account for the evolving velocity shifts 
observed in the circumstellar absorption profiles 
towards WD 1145+017. Although the model successfully accounts for the broad 
morphological changes of the line profiles, we now discuss possible alternatives and 
outstanding issues with the favored scenario.   

\subsection{Infalling/accreting material} It has been suggested that the red-shifted absorption velocities
could be caused by infalling material being accreted onto the star \citep{xu16}. 
While this scenario is realistic for the red-shifted absorption
profiles, infalling material is inconsistent with the reversal into blue-shifted absorption. 
The blue-shifted profiles, if not due to the eccentric ring model presented here, require
some sort of outflowing material, such as a disk or cometary wind. It is unlikely that we have 
observed the transition from accreting or infalling material into a wind or outflowing material, 
whereas the precessing rings explain the profile changes with no abrupt shifts in the
circumstellar material geometry.   

\subsection{High eccentricity for inner gas rings} The eccentricities required in
our model to produce the large observed velocities are in the range $0.25 - 0.30$. Similar eccentricities
have been proposed to explain the asymmetric \ion{Ca}{2} infrared triplet emission
line profiles observed in the Ton 345 system \citep{gansicke08,melis10}, while
a smaller eccentricity of $\approx 0.02$ was reported
for SDSS 1228+1040 \citep{gansicke06}. These non-zero
eccentricities have been questioned, however, and it has been suggested
instead that the asymmetric \ion{Ca}{2} line profiles, such as those
observed for SDSS J122859.93+104032.9 \citep{manser16}, may be the result of non-uniform
surface brightness in a circular gas disk \citep{metzger12}.

%Models show that 
%even small eccentricities can result in runaway evolution of gas and debris disks, quickly building up
%material at the sublimation radius and subsequently large accretion bursts \citep{metzger12}. 
%This occurs due to the strong interactions between the gas and particles
%in the debris disk.  

While non-axisymmetric disk surface brightness may be able to account for asymmetries in emission
line profiles, it cannot yield the large maximum velocities and broad velocity range
observed in the circumstellar absorption line profiles. Since the absorption arises along
a narrow line-of-sight through the material, and thermal and turbulent line broadening
cannot be greater than the local sound speed ($c_s \approx 1-2$ km s$^{-1}$), large eccentricities
are required to produce the observed line-of-sight velocities. As the eccentricity approaches zero,
the line profiles become more symmetric about zero velocity and the maximum absorption
velocities decrease. Although smaller eccentricities for smaller periastra
can produce the large absorption velocities (due to the larger Keplerian
orbital velocities at smaller orbital distances), these rings would precess far more
quickly than is observed. Thus, the circumstellar lines observed and modeled
here are consistent with the eccentric geometry suggested for the emitting gas disks
in other white dwarf systems.

It has been suggested that rocky bodies orbiting near a white dwarf's Roche radius
were initially in long period orbits at many au before being flung inward
on highly eccentric orbits by massive planets \citep[e.g.,][]{bonsor11,veras14}.
Radiation from the star can subsequently circularize the 10$^{-4} - 10$ cm sized tidally
disrupted debris in a small fraction of the white dwarf's cooling age \citep{veras15}. 
The eccentricity in the gas rings presented here could be leftover from
this circularization process. \citet{gurri17} find that any large fragments around 
WD 1145+017 are likely to be in nearly circular orbits 
if the parent body or bodies have masses greater than $10^{23}$ g. 
However, there are no such eccentricity constraints for orbiting bodies with lower 
masses and this leaves significant phase space for eccentrically orbiting 
planetesimals and their debris, consistent with the gas in our model.

Finally, it is currently unclear what the circularization timescales are for eccentric 
rings of gas around white dwarfs. If the gas viscosity is sufficiently low,
there will be little drag on the disk material and the circularization timescale may
be long. Simulations of stellar tidal disruption around supermassive black holes
show that the circularization timescale is strongly dependent on the eccentricity
of the orbiting debris and the efficiency
of energy dissipation in shocks generated by colliding material \citep[e.g.,][]{hayasaki16}. 
Smaller eccentricities, similar to the values in our model, imply lower 
relative velocities between intersecting 
gas streams and thus less energy dissipation in any resulting shocks, leading to 
longer circularization timescales. Detailed simulations of eccentric gas streams 
around white dwarfs are needed to provide constraints on the this timescale for 
scenarios similar to the favored model.

%The relative velocities between neighboring rings in the
%favored model are $\approx 50$ km s$^{-1}$, which is likely supersonic, 
%suggesting that the rings should experience
%dissipation of energy through shocks and eventually circularize, although the timescale
%is uncertain. 

%The rings in the favored model, in fact, have
%periastra near $\approx 20$ $R_*$ where silicates are expected
%to rapidly sublimate at temperatures of $\approx 1500$ K. 
%However, there is currently no direct evidence for a coincident disk of solid
%debris, although the observed transiting material is likely producing such a disk. Thus,
%although it is unclear as to the source of the ring eccentricity, some eccentricity
%is necessary to produce absorption line profiles at large line-of-sight velocities.
%\citet{gurri17} find that the orbiting debris are likely to be in nearly circular orbits 
%if the parent body or bodies have masses greater than $10^{23}$ g. 
%However, there are no such eccentricity constraints for orbiting bodies with lower 
%masses and this leaves significant phase space for eccentrically orbiting 
%planetesimals and their debris, consistent with the gas in our model.    

\subsection{An external perturber} Previous to this study, the strongest evidence of a 
precessing gas disk has been observed in SDSS 1228+1040.  In that case the data are 
consistent with a precession timescale of 24--30 years, and difficult to reconcile 
with an external, gravitational perturber such as a massive planet.  Rather, the
precession-like behavior of SDSS 1228+1040 is consistent with the effects of 
general relativity \citep{manser16}. WD 1145+017 
is inferred to have asteroid or planetesimal-sized solid bodies orbiting at $\approx 90$ $R_*$,
and thus it is possible that more massive objects, i.e., planets, are orbiting 
undetected at large distances. As such, it is not implausible
that an external perturber could be influencing secular variations in the disk.

The disk precession timescale due to external perturbing bodies can be estimated 
using dynamical simulations \citep{mustill09}. 
Assuming orbital semi-major axes of 100 $R_*$, 200 $R_*$, and
2000 $R_*$ and a perturbing mass of 1 $M_{\rm Jup}$, the resulting precession periods are
$\approx 6$, 45, and 45\,000 years. Assuming a smaller mass for the perturbing planet 
results in longer timescales. While the smallest 
precession period of $\approx 6$ years for a Jupiter-mass perturber at 100 $R_*$ is 
similar to the observed evolution timescales of the line profiles, there is no evidence
for such a massive companion at such small orbital distances. Furthermore, the Hill sphere
of a Jupiter mass planet at 100 $R_*$ is $\approx 10 R_*$ which would likely disrupt the
orbiting material near 90 $R_*$. Thus the line profile changes are almost certainly
not due to a massive perturbing body.  
      
\section{Conclusions}
\label{sec:conclusion}      

We have presented observations which reveal the complete velocity reversal of the circumstellar
line profiles towards WD 1145+017 over the course of $\approx 2$ years.      
The arguments from \autoref{sec:discussion} suggest that eccentric rings
precessing due to general relativistic effects are a reasonable
explanation for the circumstellar profile variability. The
system is highly dynamic, however, so it would not be surprising if the circumstellar
profiles continue to evolve. Additional rings with varying eccentricity can be
added to model any new features that arise in the future. Continued monitoring
of WD 1145+017 will aid in understanding the geometry of the innermost circumstllar 
gas explored here and provide constraints on white dwarf disk accretion models 
\citep[e.g.,][]{rafikov11,metzger12,kenyon17b}. 
If the eccentricity in the gas rings modeled here is intrinsic (rather
than perturbative), their origin may be collisions within eccentric
rings of solids, which is a long-expected but as-yet unconfirmed outcome
of the tidally disrupted asteroid model for polluted white dwarf systems.

\bigskip

{\bf Acknowledgments:} The authors thank the referee for their comments, which helped
improve the manuscript. The authors acknowledge informative exchanges with D. Veras and N. C. Stone. 
A portion of the data presented herein was obtained at the W.M. Keck
Observatory from telescope time allocated to the National Aeronautics and Space Administration
through the agency's scientific partnership with the California Institute of Technology and the
University of California. The Observatory was made possible by the generous financial
support of the W.M. Keck Foundation. The authors wish to recognize and acknowledge the very
significant cultural role and reverence that the summit of Mauna Kea has always had within the
indigenous Hawaiian community. We are most fortunate to have the opportunity to conduct
observations from this mountain. P. W. C. acknowledges support from the National Science
Foundation through Astronomy and Astrophysics Research Grant AST-1313268 (PI: S.R.). S. B.  
and S. R. acknowledge the National Science Foundation’s support of the Keck Northeast 
Astronomy Consortium’s REU program through grant AST-1005024. S. G. P. acknowledges the 
support of the Leverhulme Trust. The research leading to these results has received funding from the
European Research Council under the European Union's Seventh Framework
Programme (FP/2007-2013)/ERC Grant Agreement n. 320964 (WDTracer). 
This work has made use of data from ESO proposal ID 
598.C-0695. This work has made use of NASA's Astrophysics Data System.

\end{document}